\documentclass[english,prl,aps,twocolumn,10pt]{revtex4-1}

\usepackage{amsmath}
\usepackage{amssymb}
\usepackage{graphicx}
\usepackage{babel}
\usepackage{array}
\usepackage{verbatim}
\usepackage[colorlinks=true, pdfstartview=FitV, linkcolor=blue, citecolor=blue, urlcolor=blue]{hyperref} 

\def\NN        {{$^{15}$N \/}}

\def\eg       {{\it e.g.}}
\def\ie       {{\it i.e.}}

\newcolumntype{L}[1]{>{\raggedright\arraybackslash}p{#1}}
\newcolumntype{C}[1]{>{\centering\arraybackslash}p{#1}}

\newcommand{\captionstyle}{\normalfont} 

\newcommand{\mr}[1]{\mathrm{#1}}
\newcommand{\unit}[1]{\,\mathrm{#1}}

\newcommand{\us}{\,\mu{\rm s}}
\newcommand{\uT}{\,\mu{\rm T}}

\newcommand{\rtHz}{\sqrt{\mr{Hz}}}

\newcommand{\ye}{\gamma_\mr{e}}

\newcommand{\Bmin}{B_\mr{min}}
\newcommand{\Dpmax}{\Delta p_\mr{max}}
\newcommand{\Dphimax}{\Delta\phi_\mr{max}}

\newcommand{\tint}{t_\mr{int}}
\newcommand{\tm}{t_\mr{m}}
\newcommand{\tpi}{t_\pi}
\newcommand{\trep}{t_\mr{rep}}
\newcommand{\ts}{t_s}
\newcommand{\tw}{t_\mr{rep}}
\newcommand{\wo}{\omega_0}

\begin{document}

\title{Reconstruction-free quantum sensing of arbitrary waveforms}

\author{J. Zopes and C. L. Degen$^1$}
\affiliation{$^1$Department of Physics, ETH Zurich, Otto Stern Weg 1, 8093 Zurich, Switzerland.}
\email{degenc@ethz.ch}

\begin{abstract}
We present a protocol for directly detecting time-dependent magnetic field waveforms with a quantum two-level system.  Our method is based on a differential refocusing of segments of the waveform using spin echoes.  The sequence can be repeated to increase the sensitivity to small signals.  The frequency bandwidth is intrinsically limited by the duration of the refocusing pulses.  We demonstrate detection of arbitrary waveforms with $\sim 20\unit{ns}$ time resolution and $\sim 4\unit{\uT/\rtHz}$ field sensitivity using the electronic spin of a single nitrogen-vacancy center in diamond.
\end{abstract}

\date{\today}

\maketitle


Well-controlled two-level quantum systems with long coherence times have proven useful for precision sensing \cite{budker07,degen17} of various physical quantities including temperature \cite{kucsko13}, pressure \cite{doherty13}, or electric \cite{dolde11} and magnetic fields \cite{loretz13,zopes17}.  By devising suitable coherent control sequences, such as dynamical decoupling \cite{delange10}, quantum sensing has been extended to time-varying signals.  In particular, coherent control schemes have allowed the recording of frequency spectra \cite{bylander11,schmitt17,boss17} and lock-in measurements of harmonic test signals \cite{kotler11}.

A more general task is the recording of arbitrary waveform signals, in analogy to the oscilloscope in electronic test and measurement.  In this case, conventional dynamical decoupling sequences are no longer the method of choice as the sensor output is non-trivially connected to the input waveform signal, requiring alternative sensing approaches.
For slowly varying signals, the transition frequency of the sensor can be tracked in real time \cite{schoenfeld11}, permitting detection of arbitrary waveforms in a single shot.  By using a large ensemble of quantum sensors detection bandwidths of up to $\sim 1\unit{MHz}$ have been demonstrated \cite{dezanche08,shin12},
with applications in MRI tomograph stabilization \cite{dezanche08}, neural signaling \cite{jensen16,barry16}, or magnetoencephalography \cite{xia06}.

For rapidly changing signals the waveform can no longer be tracked, and a general waveform cannot be recorded in a single shot.  However, if a waveform is repetitive or can be re-triggered, multiple passages of the waveform can be combined to reconstruct the full waveform signal.  This method, known as equivalent-time sampling, is routinely implemented in digital oscilloscopes to capture signals at effective sampling rates that are much higher than the rate of analog-to-digital conversion.
In quantum sensing, one possibility is to record a series of time-resolved spectra that cover the duration of the waveform \cite{zopes18prl}.  This method, however, is limited to strong signals because the spectral resolution inversely scales with the time resolution.  Other approaches include pulsed Ramsey detection \cite{balasubramanian09}, Walsh dynamical decoupling \cite{magesan13,cooper14}, and Haar wavelet sampling \cite{Xu16}, discussed below.  These methods use coherent control of the sensor to achieve competitive sensitivities, but require some form of waveform reconstruction.


In this Letter we experimentally demonstrate a simple quantum sensing sequence for directly recording time-dependent magnetic fields with no need for signal reconstruction.  Our method uses a spin echo to differentially detect short segments of the waveform, and achieves simultaneous high magnetic field sensitivity and high time resolution.  The only constraints are that the waveform can be triggered twice within the coherence time of the sensor, and that the signal amplitude remains within the excitation bandwidth of qubit control pulses.  
Possible applications include the \textit{in situ} calibration of miniature radio-frequency transmitters \cite{sasaki18,zopes18prl}, activity mapping in integrated circuits \cite{nowodzinski15}, detection of pulsed photocurrents \cite{zhou19}, and magnetic switching in thin films \cite{baumgartner17}.

\begin{figure}[t]
\includegraphics[width=1.0\columnwidth]{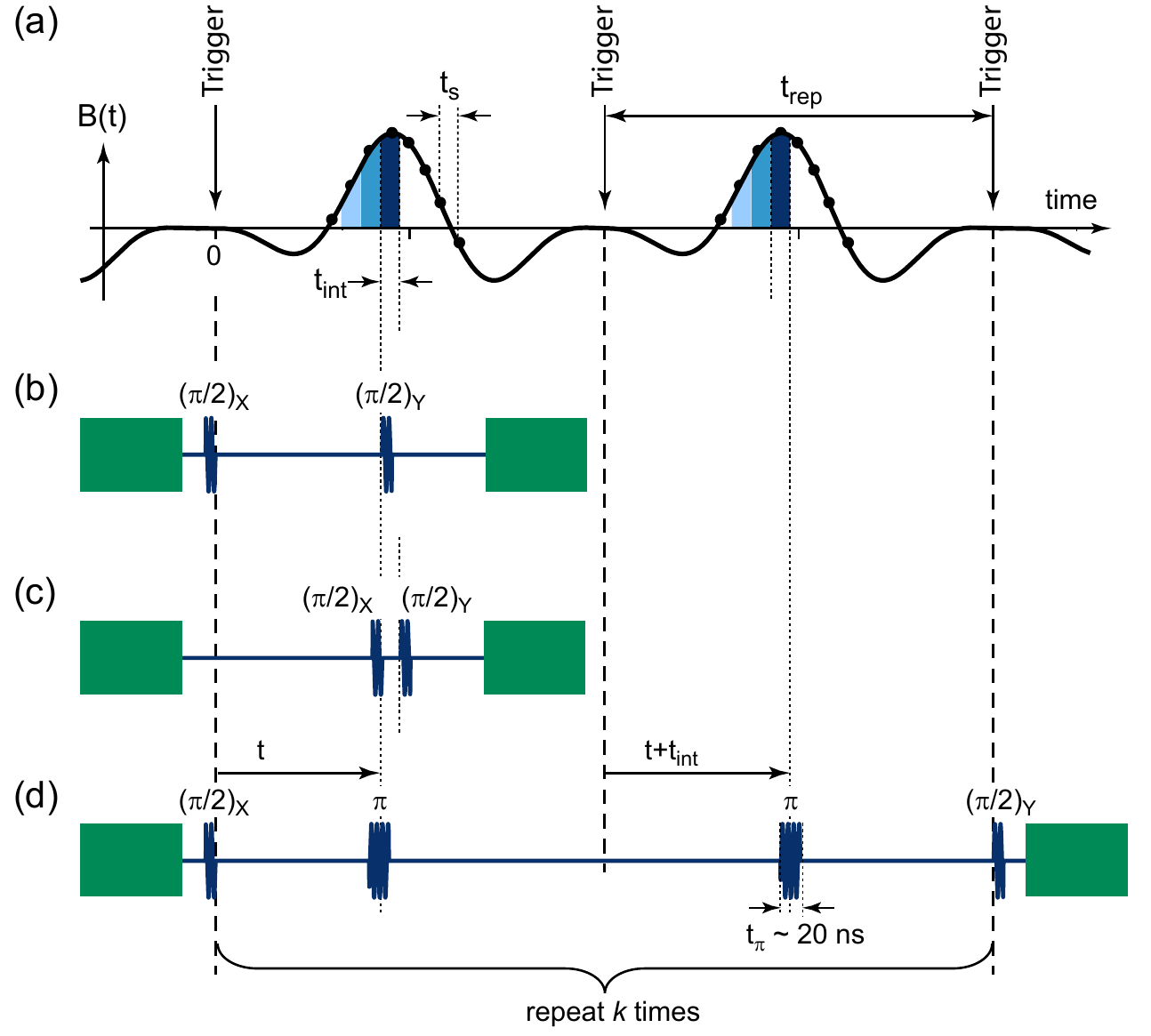}
\caption{\captionstyle
\textbf{Schemes for equivalent-time waveform sampling by a quantum sensor.}
(a) Schematic of a repetitive arbitrary waveform $B(t)$.  $t$ is the time relative to the preceding trigger and $\trep$ is the repetition time.
Dots indicate the waveform sampling and $\ts$ is the sampling time.
(b) Standard integrative Ramsey detection of the waveform.  The sensor acquired phase is proportional to the integral of the waveform between $0$ and $t$.  Signals are detected by stepping $t$ in increments of $\ts$.  Microwave pulses are shown in dark blue. Labels indicate the pulse angles and subscripts the pulse phases.
Laser arm and readout pulses are shown in green.
(c) Small interval Ramsey detection of the waveform.  
(d) Differential detection of the waveform by spin echoes.  $\tint$ is the differential integration time and $\tpi$ is the $\pi$ pulse duration.  The differential protocol can be repeated $k$ times to linearly increase the accumulated phase.
\label{fig:fig1}
}
\end{figure}
%


To motivate our measurement protocol we first inspect the interferometric Ramsey method, which has been a standard method for early quantum sensing of waveforms \cite{balasubramanian09}.  In a Ramsey experiment a superposition state, prepared by a first $\pi/2$ pulse, evolves during a sensing time $t$ and acquires a phase factor $\phi(t)$ that is proportional to the transition frequency $\wo$ between ground and excited states (see Fig. \ref{fig:fig1}(b)).  For a spin sensor, where $\wo$ is proportional to the component of the magnetic field along the spin's quantization axis, the acquired phase is
\begin{align}
\phi(t) = \int_{0}^{t} \ye B(t') dt'. 
\label{eq:ramsey}
\end{align}
Here, $B(t)$ is the time-dependent magnetic field that we aim to measure and $\ye$ is the gyromagnetic ratio of the spin.  To extract the phase, $\phi(t)$ is typically converted into a population difference $p(t)$ by a second $\pi/2$ pulse,
\begin{align}
p(t) = \frac{1}{2} (1+\sin(\phi(t))) \overset{\phi \ll 1}{\approx} \frac{1}{2} (1+\phi(t)).
\label{eq:linsensor}
\end{align}
followed by a projective readout of the sensor and signal averaging \cite{degen17}.  By measuring $p(t)$ as a function of $t$, one thus effectively measures the integral of the magnetic field in the interval $[0,t]$.  Using a numerical derivative the magnetic field can subsequently be reconstructed \cite{balasubramanian09}. However, this reconstruction greatly increases noise due to the derivative \cite{knowles14} and often requires phase unwrapping.

A more direct method that avoids numerical processing is the sampling of the waveform in small intervals $\tint$ and to build up the waveform by stepping $t$.  The simplest approach is use a Ramsey sequence with a very short integration time $\tint$ (Fig. \ref{fig:fig1}(c)).  In this case the sensor phase $\phi(t)$ encodes the field in the time interval $[t,t+\tint]$,
\begin{align}
\phi(t) = \int_{t}^{t+\tint} \ye B(t') dt' \approx \ye B(t) \tint \ ,
\end{align}
without the need for numerical post-processing.  Thanks to the short $\tint$ one can often take advantage of the linear approximation ($\sin\phi\approx \phi$) in Eq. (\ref{eq:linsensor}).  The short $\tint$, however, impairs sensitivity because $\phi \propto \tint$.

To maintain adequate sensitivity even for short $\tint$ we introduce a detection protocol that accumulates phase from several consecutive waveform passages.  Our scheme requires that the repetition time is short, $\trep\ll T_2$, where $T_2$ is the sensor's coherence time, which is often the case for fast waveform signals.  Our protocol is shown in Fig. \ref{fig:fig1}(d): By inserting two $\pi$ pulses at times $t$ and $t+\tint$ relative to two consecutive waveform triggers, we selectively acquire phase from the time interval $[t,t+\tint]$ while canceling all other phase accumulation.
A similar scheme of partial phase cancellation has been implemented with digital Walsh filters \cite{cooper14} and Haar functions \cite{Xu16} via a sequence of $\pi$ rotations.
The linear recombination of sensor outputs in such waveform sampling, however, is prone to introducing errors, especially for rapidly varying signals whose detection requires many $\pi$ pulses \cite{magesan13}.
In our scheme, the $\pi$ rotations effectively act as an \textit{in situ} derivative to the phase integral (Eq. \ref{eq:ramsey}), bypassing the need for a later numerical differentiation or reconstruction.  To further amplify the signal, the basic two-$\pi$-pulse block can be repeated $k$ times to accumulate phase from $2k$ waveform passages, up to a limit set by $2k\trep \leq T_2$.  The amplified signal is (in linear approximation)
\begin{align}
p(t) \approx 0.5 + 2 k\ye B(t) \tint \ ,
\label{eq:phik}
\end{align}
and when converted to units of magnetic field, 
\begin{align}
B(t) \approx \frac{p(t)-0.5}{ 2 k \ye \tint} \ .
\end{align}
\begin{figure}[t]
\includegraphics[width=1.0\columnwidth]{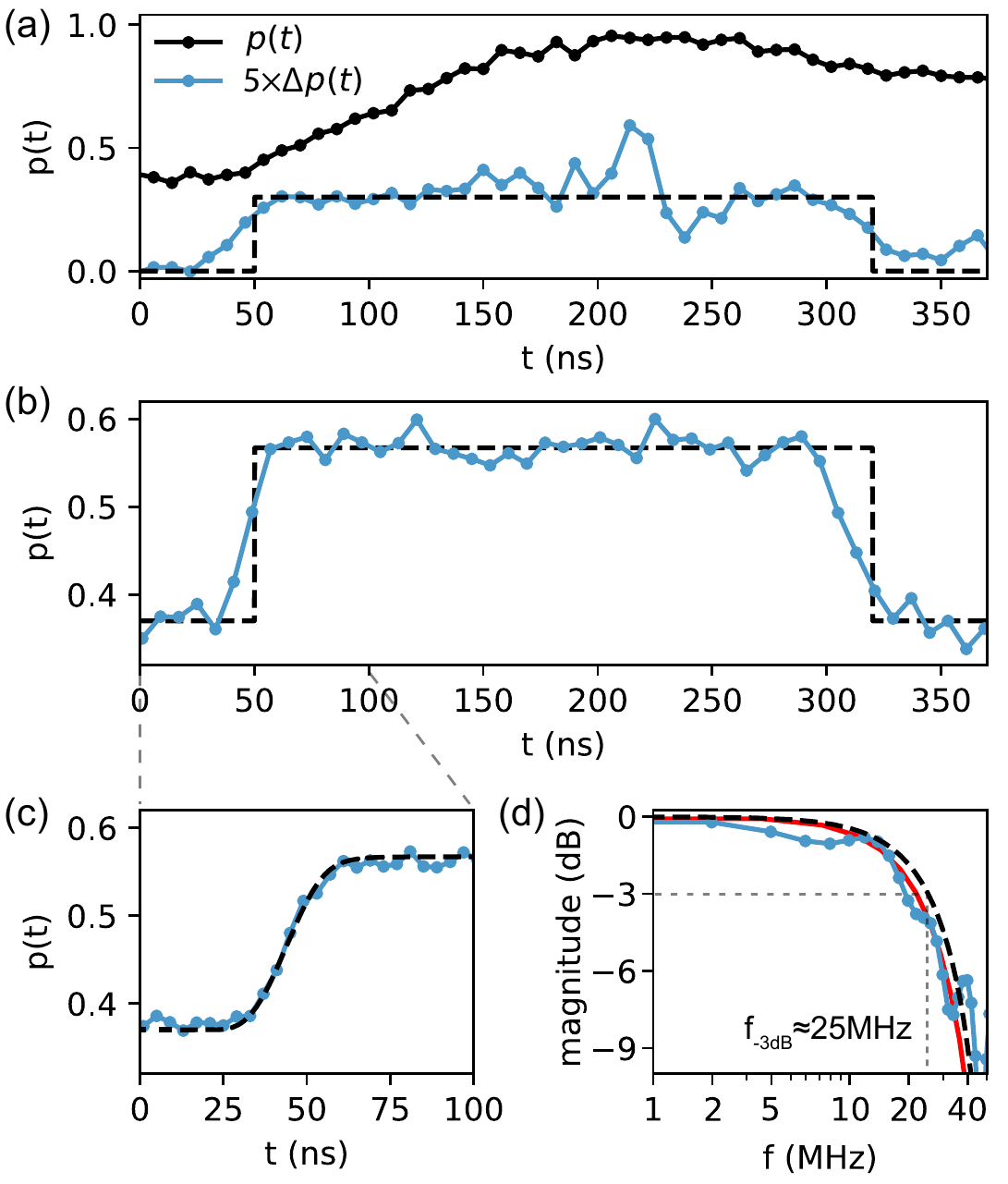}
\caption{\captionstyle
\textbf{Pulse detection and time resolution.}
(a) Sensor response to a square-wave magnetic signal (dashed curve) recorded using the standard integrative Ramsey method [protocol of Fig. \ref{fig:fig1}(b)].  The raw data are shown in black and the reconstructed waveform is shown in blue.  The waveform is reconstructed by applying a 4-point moving average to the raw data and calculating the difference $\Delta p$ between adjacent points.  The dwell time is $t_s=8\,\mathrm{ns}$ and the total averaging time is 1 hour.
(b) Sensor response (raw signal) to the same waveform signal recorded using the differential spin echo technique [protocol of Fig. \ref{fig:fig1}(d) using $k=2$].  The total averaging time is 15 min.
(c) High resolution sampling ($t_s=4\,\mathrm{ns}$) of the rising edge of the square pulse waveform.  The blue points are measured data.  The dashed black line is the expected step response for $\pi$-pulse and integration lengths of $\tint=\tpi=20\unit{ns}$.
(d) Magnitude plot of the corresponding sensor transfer function.  Blue dots are the data and the black dashed curve is the Fourier transform of a Hann window of duration $2\tpi=40\unit{ns}$.  The red curve additionally takes the finite response time of the test signal circuit ($\sim 8\unit{ns}$) into account.
\label{fig:fig2}
}
\end{figure}
\begin{figure*}[t]
\includegraphics[width=1.0\textwidth]{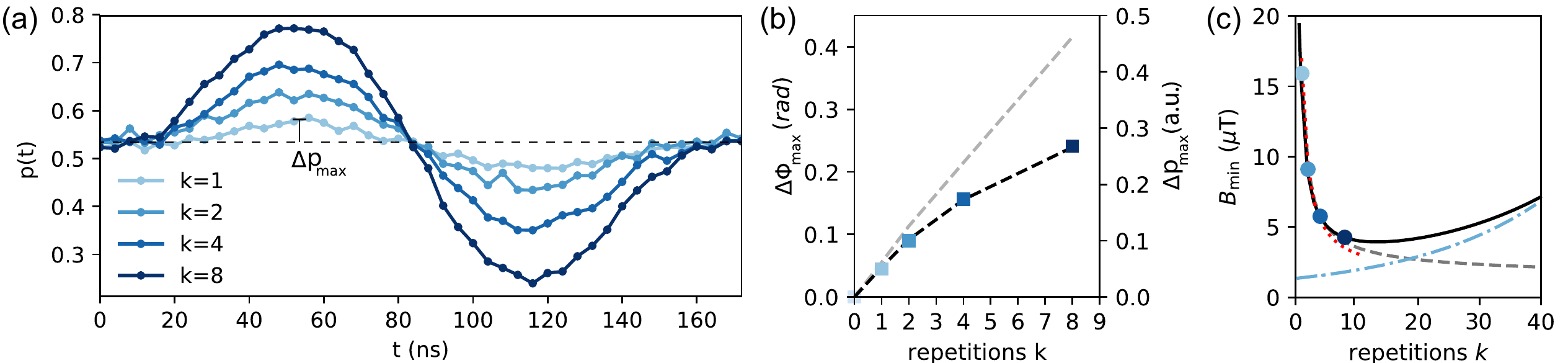}
\caption{\captionstyle
\textbf{Increased sensitivity by integrating $2k$ waveform passages.}
(a) Sensor output $p(t)$ for $k=1,2,4$ and 8 repetitions of the two-$\pi$-pulse unit (see Fig. \ref{fig:fig1}(c)), for a sine waveform of amplitude $10\unit{\uT}$ and frequency $f=4\unit{MHz}$.  The integration time and $\pi$-pulse duration are $\tint=\tpi=20\unit{ns}$ and the repetition time is $\trep=344\unit{ns}$.
(b) Peak output signal $\Dpmax$ as a function of $k$ (colored squares).  The gray dashed line shows a linear scaling that would be expected in the absence of sensor decoherence.  The black dash-dotted line takes decoherence into account ($T_2 = 14\unit{\us}$).
(c) Minimum detectable magnetic field $\Bmin$ per unit time as defined by Eq. (\ref{eq:Bmin}) (black curve).  Colored dots represent the data from (a).  The dashed, dash-dotted and dotted curves are explained in the text.
\label{fig:fig3}
}
\end{figure*}
%


We experimentally demonstrate arbitrary waveform sampling using the electronic spin of a single nitrogen-vacancy (NV) center in a diamond single crystal.  The NV spin is initialized and read out using $\sim 2\unit{\us}$ green laser pulses and a single-photon-counting module \cite{loretz13}.  Microwave control pulses are generated by an arbitrary waveform generator (AWG), amplified to reach Rabi frequencies of $\sim 25\unit{MHz}$, and applied to the NV center via a coplanar waveguide (CPW) structure \cite{zopes17}. 
Test magnetic waveforms are generated by a second function generator operated in burst mode and triggered by the AWG.  The test signals are delivered to the NV center either by injecting them into the common CPW using a bias-T \cite{rosskopf17} or by an auxiliary nearby microcoil \cite{zopes18prl,zopes18ncomms}.
The setup is operated in a magnetic bias field of $43\unit{mT}$ (aligned with the N-V crystal direction) to isolate the $\{m_s=0,m_s=-1\}$ manifold of the $S=1$ NV spin, and to achieve preferential alignment of the intrinsic nitrogen nuclear spin (here the spin 1/2 of the \NN isotope) \cite{jacques09}.  The latter is not required for our scheme, but helps reducing microwave pulse errors.


We begin our study by recording a simple, 270-ns-long square waveform (Fig. \ref{fig:fig2}).  We record the waveform both using the standard integrative Ramsey scheme [Fig. \ref{fig:fig1}(b)] and our differential sampling technique [Fig. \ref{fig:fig1}(d)].
For the Ramsey scheme, we reconstruct the magnetic waveform by a numerical differentiation of the raw signal (black data in Fig. \ref{fig:fig2}(a)) via the central difference quotient of the smoothed signal \cite{jordan17}.  The reconstructed waveform is shown in blue.
For our differential detection scheme, we directly plot the signal output without any further data processing (Fig. \ref{fig:fig2}(b)).  Clearly, the differential sampling method is able to faithfully reproduce the square pulse and is not affected by the noise amplification of the Ramsey scheme.

To characterize the time resolution of the method, we record the rising edge of the pulse with fine sampling $\ts = 4\unit{ns}$ (Fig. \ref{fig:fig2}(c)).  We find a 10-90\% step response time of $\tau \sim 20\unit{ns}$.  The response time is approximately given by $\tau \approx \max(\tpi,\tint)$, since the finite pulse duration and the integration time both act as moving average filters.  While $\tint$ can be deliberately adjusted, $\tpi$ is determined by the Rabi frequency of the system and sets a hard limit to the response time.

In Fig. \ref{fig:fig2}(d) we show the corresponding frequency transfer function $G(\omega)$ of the sensor, \ie, the Fourier transform of the unit impulse response obtained from the step response. In our experiments, where $\tint = \tpi$, the unit impulse response of the sensor is approximately given by a Hann function with characteristic length $2\tpi$ \cite{supplementary}.  The Bode plot indicates a -3dB sensor bandwidth $f_{-3\mathrm{dB}} \approx 25\unit{MHz}$, with good agreement between theory and experiments.
This bandwidth could be slightly increased, up to $\sim 40\unit{MHz}$ \cite{supplementary}, by choosing shorter integration times $\tint\ll\tpi$; however, the short integration time comes with the penalty of vanishing sensitivity.

In a next step, we investigate the signal gain possible by accumulating phase from $2k$ consecutive waveform passages.  Fig. \ref{fig:fig3}(a) plots the sensor response from a weak sinusoidal test signal recorded with $k=1,2,4$ and $8$.  Clearly, a much larger oscilloscope response results for higher $k$ values.
To estimate the signal gain, we plot the peak sensor signal $\Dpmax$ (indicated in (a)) as a function of $k$, see Fig. \ref{fig:fig3}(b).  At small $k$ values the increase of $\Dpmax$ is proportional to $k$, as expected, while at larger $k$ decoherence of the sensor attenuates the signal.  By correcting for sensor decoherence, we can recover the almost exact linear scaling of the signal phase $\Dphimax$ with $k$ (dashed line in (b)).

To quantify the overall sensitivity in the presence of decoherence and sensor readout overhead, we calculate a minimum detectable field $\Bmin$, defined as the input field that gives unity signal-to-noise ratio for a one-second integration time. $\Bmin$ is given by \cite{degen17}, 
\begin{align}
\Bmin = \frac{\sqrt{\tm+2 k \tw} e^{\frac{2 k \tw}{T_2}}}{2 \ye k C \tint} \ ,
\label{eq:Bmin}
\end{align}	
where $\tm = 3\unit{\us}$ is the arm/readout duration (see Fig. \ref{fig:fig1}(c)), $T_2\sim 14\unit{\us}$ is the coherence time, and $C\sim 0.02$ is a dimensionless number that quantifies the quantum readout efficiency \cite{degen17}. 
In Fig. \ref{fig:fig3}(c) we plot $\Bmin$ as a function of $k$.  We find that $\Bmin \propto k^{-1}$ for short durations $k\tw < \tm$, that is, the benefit of repeating the sequence is largest for small $k$ and high repetition rates (dotted curve).  Once $k\tw > \tm$ the scaling reduces to $\Bmin \propto k^{-0.5}$ because the linear phase accumulation now competes with standard signal averaging (dashed curve).  For large $k\tw$ that exceed the sensor coherence time $T_2$ the efficiency of the method rapidly deteriorates (dash-dotted curve).

\begin{figure}[t]
\includegraphics[width=1.0\columnwidth]{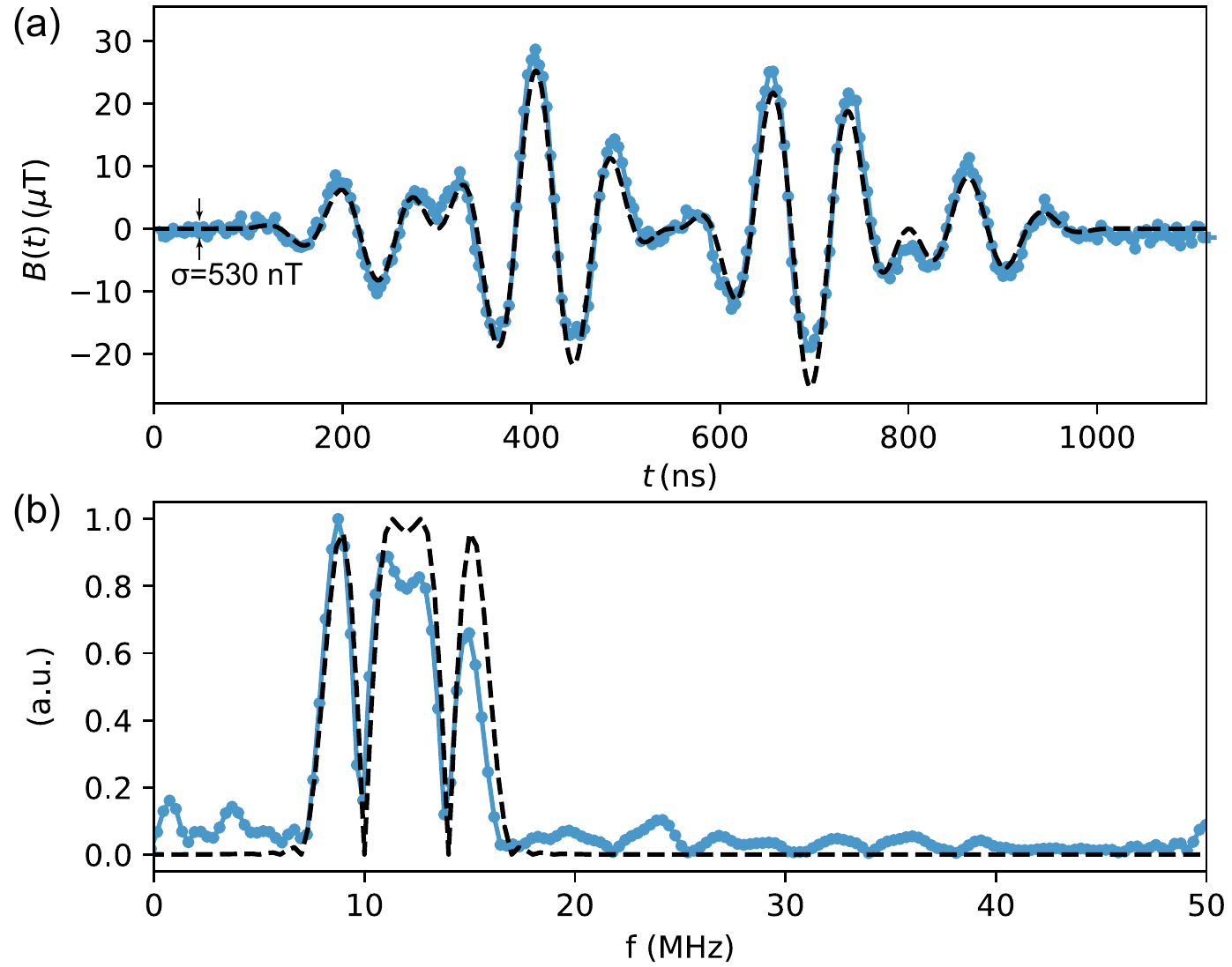}
\caption{\captionstyle \textbf{Example of arbitrary waveform detection.}
(a) Input waveform (dashed line) and recorded waveform data (blue dots) for a complex waveform given by 
$B(t) = B \sin^2 \left(\omega t/2 \right) \left[ \sin \left(12 \omega t \right) \cos \left(\omega t \right) \sin^2 \left(\omega t \right) \right]$, with $B = 81.87\unit{\uT}$ and $\omega=2\pi \times 1\unit{MHz}$.  The waveform is sampled using $N=280$ data points and $\ts = 4\unit{ns}$.  Further parameters are $\tint=\tpi=20\unit{ns}$, $\trep = 1400\unit{ns}$ and $k=4$.  The total experimental time is $60\unit{h}$ corresponding to $\sim 1.44\times 10^{10}$ waveform triggers. The baseline noise is $\sigma\approx 530\,$nT-rms.
(b) Normalized power spectra of the input waveform (black dashed line) and the detected waveform (light blue connected points). Fourier components at higher frequencies are slightly attenuated due to the limited bandwidth of the sensing sequence.
\label{fig:fig4}
}
\end{figure}

We complete our study by demonstrating detection of a complex test waveform (Fig. \ref{fig:fig4}).  The waveform contains the sum of several Fourier components with the analytical expression for $B(t)$ given in the figure caption.  In Fig. \ref{fig:fig4}(a) we show the experimentally measured waveform (light blue points) together with the input waveform (dashed black line) in the same plot.  The experimental waveform consists of $N=280$ data points sampled at $\ts = 4\unit{ns}$ horizontal resolution.  Clearly, the experimental waveform agrees very well with the applied input.  The experimental data are plotted without any data processing, demonstrating that our differential sampling method directly reproduces the waveform signal.  Fig. \ref{fig:fig4} (b) further presents the corresponding power spectra of the input waveform (black dashed line) and the recorded sensor output (light blue points).  Although the signal lies within the analog bandwidth of the sensor ($\sim 25\unit{MHz}$), some attenuation is observed at higher frequencies.  If desired, inverse filtering techniques could be applied to compensate the high-frequency roll-off of the sensor.

Before concluding, we point out a few limitations and possible remedies of the differential waveform sampling technique.
First, our scheme is only applicable to waveforms that can be triggered twice within the sensors $T_2$ time.  While $T_2$ could be extended to some extent by adding dynamical decoupling $\pi$ pulses to our protocol, very long repetition times cannot be covered, and will require resorting to, \eg, the inefficient small-interval Ramsey technique (Fig. \ref{fig:fig1}(c)).
Second, the maximum peak-to-peak signal amplitude is limited by the excitation bandwidth of $\pi$ pulses to $(\ye\tpi)^{-1}$, here $\sim 2\unit{mT}$.  Only relatively weak fields can therefore be detected with our method.  To cover strong signals, time-resolved spectroscopy techniques are available \cite{zopes18prl}.
Third, when accumulating signal over many passages $k$, the phase may exceed the sensor's linear range (see Eq. \ref{eq:ramsey}).  In this situation, the relative phase of the second $\pi/2$ pulse could be cycled \cite{knowles16} to recover a linear response.


In summary, we have presented a quantum sensing method for direct detection of arbitrary waveforms in the time domain using equivalent time sampling.  Our method does not require any waveform reconstruction, allowing, for example, to sample arbitrary segments from a longer waveform.  In addition, our protocol can be repeated to coherently accumulate phase from many waveform cycles to improve sensitivity.  The analog bandwidth of our scheme is fundamentally limited by the Rabi frequency of the sensor, which sets the minimum $\pi$ pulse duration $\tpi$.  In the present work, we demonstrate a time resolution of $\tpi \sim 20\unit{ns}$ using a Rabi frequency of $\sim 25\unit{MHz}$.  To achieve better time resolution, the Rabi frequency could be increased by more than an one order of magnitude by miniaturizing the coplanar waveguide \cite{fuchs09,kong18}.  The highest demonstrated Rabi frequencies are $200-500\unit{MHz}$ for NV centers, corresponding to $\tpi = 1-2.5\unit{ns}$ \cite{fuchs09,kong18}.  At this time resolution it may become feasible to study the photoresponse in materials \cite{zhou19} or the switching in thin film magnetic memory devices \cite{baumgartner17}.


We thank Pol Welter, Martin W\"ornle and Konstantin Herb for helpful discussions.
This work has been supported by Swiss National Science Foundation (SNFS) Project Grant No. 200020\_175600, the National Center of Competence in Research in Quantum Science and Technology (NCCR QSIT), and the Advancing Science and TEchnology thRough dIamond Quantum Sensing (ASTERQIS) program, Grant No. 820394, of the European Commission.

%

\end{document}